\documentclass[sigconf,screen]{acmart} 

\usepackage{mathrsfs}
\usepackage{subfigure}
\usepackage{microtype,flushend}

\usepackage{balance}

\setcopyright{acmcopyright}
\acmPrice{15.00}
\acmDOI{10.1145/3368089.3409759}
\acmYear{2020}
\copyrightyear{2020}
\acmSubmissionID{fse20main-p719-p}
\acmISBN{978-1-4503-7043-1/20/11}
\acmConference[ESEC/FSE '20]{Proceedings of the 28th ACM Joint European Software Engineering Conference and Symposium on the Foundations of Software Engineering}{November 8--13, 2020}{Virtual Event, USA}
\acmBooktitle{Proceedings of the 28th ACM Joint European Software Engineering Conference and Symposium on the Foundations of Software Engineering (ESEC/FSE '20), November 8--13, 2020, Virtual Event, USA}

\begin{document}

\title{A Comprehensive Study on Challenges in Deploying Deep Learning Based Software}

\author{Zhenpeng Chen}
\affiliation{%
  \institution{Key Lab of High-Confidence Software Technology, MoE (Peking University)}
  \country{Beijing, China}
}
\email{czp@pku.edu.cn}

\author{Yanbin Cao}
\affiliation{%
  \institution{Key Lab of High-Confidence Software Technology, MoE (Peking University)}
  \country{Beijing, China}
}
\email{caoyanbin@pku.edu.cn}

\author{Yuanqiang Liu}
\affiliation{%
  \institution{Key Lab of High-Confidence Software Technology, MoE (Peking University)}
  \country{Beijing, China}
}
\email{yuanqiangliu@pku.edu.cn}

\author{Haoyu Wang}
\affiliation{%
  \institution{Beijing University of Posts and Telecommunications}
  \country{Beijing, China}
}
\email{haoyuwang@bupt.edu.cn}

\author{Tao Xie}
\affiliation{%
  \institution{Key Lab of High-Confidence Software Technology, MoE (Peking University)}
  \country{Beijing, China}
}
\email{taoxie@pku.edu.cn}

\author{Xuanzhe Liu}\authornote{Corresponding author: Xuanzhe Liu (xzl@pku.edu.cn).}
\affiliation{%
  \institution{Key Lab of High-Confidence Software Technology, MoE (Peking University)}
  \country{Beijing, China}
}
\email{xzl@pku.edu.cn}


\newcommand{\para}[1]{\smallskip\noindent{\bf {#1}. }}
\newcommand{\czp}[1]{{\color{red}{#1}}}
\newcommand{\yanbin}[1]{{\color{blue}{lx: #1}}}

\begin{abstract}
Deep learning (DL) becomes increasingly pervasive, being used in a wide range of software applications. These software applications, named as DL based software (in short as \emph{DL software}), integrate DL models trained using a large data corpus with DL programs written based on DL frameworks such as TensorFlow and Keras. A DL program encodes the network structure of a desirable DL model and the process by which the model is trained using the training data. 
To help developers of DL software meet the new challenges posed by DL, enormous research efforts in software engineering have been devoted. Existing studies focus on the development of DL software and extensively analyze faults in DL programs. However, the deployment of DL software has not been comprehensively studied. To fill this knowledge gap, this paper presents a comprehensive study on understanding challenges in deploying DL software. We mine and analyze 3,023 relevant posts from Stack Overflow, a popular Q\&A website for developers, and show the increasing popularity and high difficulty of DL software deployment among developers. We build a taxonomy of specific challenges encountered by developers in the process of DL software deployment through manual inspection of 769 sampled posts and report a series of actionable implications for researchers, developers, and DL framework vendors.

\end{abstract}



\maketitle

\section{Introduction}\label{intro}
Deep learning (DL) has been used in a wide range of software applications from different domains, including natural language processing~\cite{wwwChenSHLML19}, speech recognition~\cite{aaaiMcMahanR18}, image processing~\cite{aaaiFurutaIY19}, disease diagnosis~\cite{icdmGaoWWYGWT019}, autonomous driving~\cite{cvpr0001CS19}, etc. 
These software applications, named as DL based software (in short as \emph{DL software}), integrate DL models trained using a large data corpus with DL programs. To implement DL programs, developers rely on DL frameworks (e.g., TensorFlow~\cite{osdiAbadiBCCDDDGIIK16} and Keras~\cite{keraslink}), which encode the structure of desirable DL models and the process by which the models are trained using the training data. 

The increasing dependence of current software applications on DL (as in DL software) makes it a crucial topic in the software engineering (SE) research community.
Specifically, many research efforts~\cite{issredeep19,isstaZhangCCXZ18,sigsoftIslamNPR19,corrabstaxonomy,corrabseduc} have been devoted to characterizing the new challenges that DL poses to \emph{software development}. 
To characterize the challenges that developers encounter in this process, various studies~\cite{issredeep19,isstaZhangCCXZ18,sigsoftIslamNPR19} focus on analyzing faults in DL programs. For instance, Islam et al.~\cite{sigsoftIslamNPR19} have presented a comprehensive study of faults in DL programs written based on TensorFlow (TF)~\cite{osdiAbadiBCCDDDGIIK16},  Keras~\cite{keraslink},  PyTorch~\cite{pytorchlink}, Theano~\cite{RfouAAa16}, and Caffe~\cite{JiaSDKLGGD14} frameworks.

Recently, the great demand of deploying DL software to different platforms for real usage~\cite{corrabseduc,wwwMaXZTL19,wwwXuLLLLL19} also poses new challenges to \emph{software deployment}, i.e., deploying  DL software on a specific platform. For example, a computation-intensive DL model in DL software can be executed efficiently on PC platforms with the GPU support, but it cannot be directly deployed and executed on platforms with limited computing power, such as mobile devices. To facilitate this deployment process, some DL frameworks such as TF Lite~\cite{tflitelink} and Core ML~\cite{coremllink} are rolled out by major vendors. Furthermore, SE researchers and practitioners also begin to focus on DL software deployment. For example, Guo et al.~\cite{GuoCXMHLLZL19} have investigated the changes in prediction accuracy and performance when DL models trained on PC platforms are deployed to mobile devices and browsers, and unveiled that the deployment still suffers from compatibility and reliability issues. Additionally, DL software deployment also poses some specific programming challenges to developers such as converting DL models to the formats expected by the deployment platforms; these challenges are frequently asked in developers' Q\&A forums~\cite{so61,so23,so25,so26}.
Despite some efforts made, to the best of our knowledge, a fundamental question remains under-investigated: \emph{what specific challenges do developers face when deploying DL software}?

To bridge the knowledge gap, this paper presents the first comprehensive empirical study on identifying challenges in deploying DL software. Given surging interest in DL and the importance of DL software deployment, this study can aid developers to avoid common pitfalls and make researchers and DL framework vendors\footnote{Unless explicitly stated, framework vendors in this paper refer to vendors of deployment related frameworks such as TF Lite and Core ML.} better positioned to help software engineers  perform the deployment task in a more targeted way. Besides mobile devices and browsers that have been considered in previous work~\cite{GuoCXMHLLZL19}, in this work, we also take into account server/cloud platforms, where a large number of DL software applications are deployed~\cite{corrabs191003156,corrabseduc}. To understand what struggles that developers face when they deploy DL software, we analyze the relevant posts from a variety of developers on Stack Overflow (SO), which is one of the most popular Q\&A forums for developers. When developers have troubles in solving programming issues that they meet, they often seek technological advice from peers on SO~\cite{corrabseduc}. Therefore, it has been a common practice for researchers to understand the challenges that developers encounter when dealing with different engineering tasks from SO posts, as shown in recent work~\cite{corrabseduc,msrBajajPM14,AlshangitiSMLY19,issredeep19,sigsoftBagherzadehK19,esemAhmedB18,jcstYangLXWS16,eseRosenS16}.

Our study collects and analyzes 3,023 SO posts regarding deploying DL software to server/cloud, mobile, and browser platforms. Based on these posts, we focus our study on the following research questions.

\textbf{RQ1: Popularity trend.} Through quantitative analysis, we find that DL software deployment is gaining increasing attention, and find evidence about the timeliness and urgency of our study.

\textbf{RQ2: Difficulty.} We measure the difficulty of DL software deployment using well-adopted metrics in SE. Results show that the deployment of DL software is more challenging compared to other aspects related to DL software. This finding motivates us to further unveil the specific challenges encountered by developers in DL software deployment.

\textbf{RQ3: Taxonomy of challenges.} To identify specific challenges in DL software deployment, we randomly sample a set of 769 relevant SO posts for manual examination. For each post, we qualitatively extract the challenge behind it. Finally, we build a comprehensive taxonomy consisting of 72 categories, linked to challenges in deploying DL software to server/cloud, mobile, and browser platforms. The resulting taxonomy indicates that DL software deployment faces a wide spectrum of challenges.

Furthermore, we discuss actionable implications (derived from our results) for researchers, developers, and DL framework vendors. In addition, this paper offers a dataset of posts related to DL software deployment~\cite{smater} as an additional contribution to the research community for other researchers to replicate and build upon.

\section{Background}\label{back}
\begin{figure}
\includegraphics[width=1.0\columnwidth]{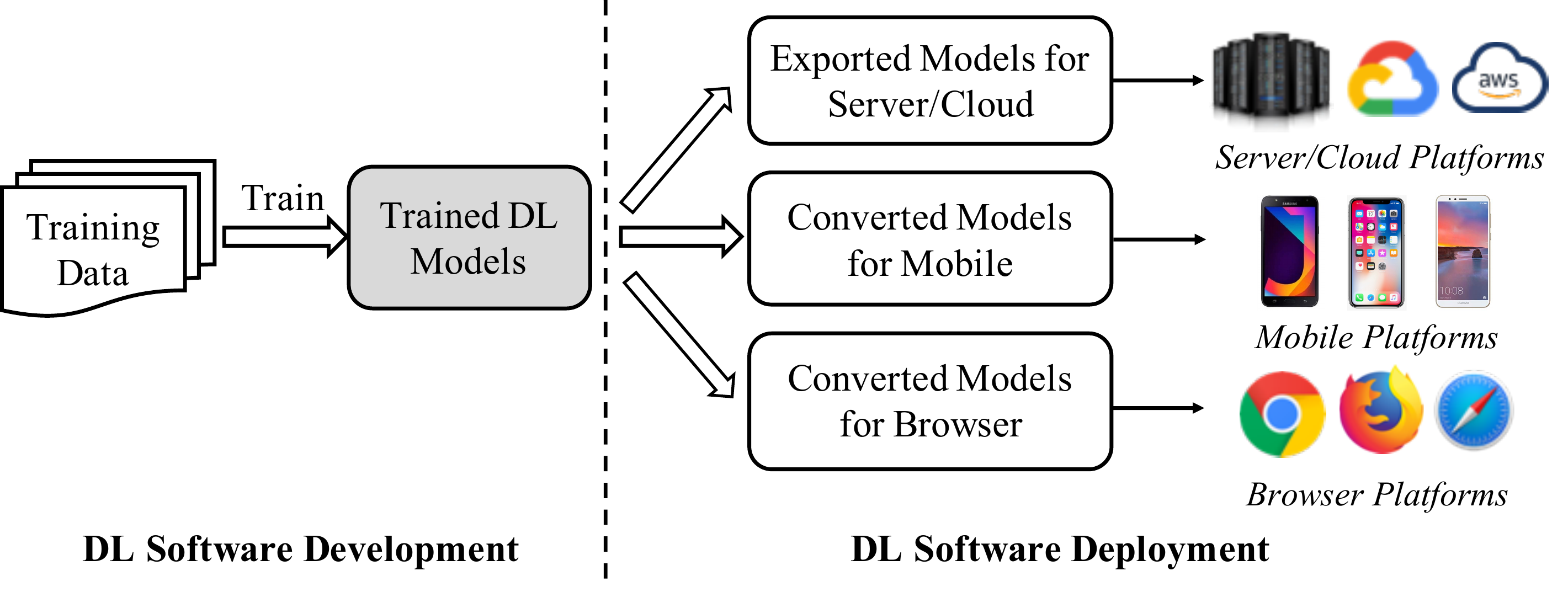}
\caption{DL software development and deployment.}\label{fig:devordep}
\end{figure}

We first briefly describe the current practice of DL software development and deployment. Figure~\ref{fig:devordep} distinguishes the two processes. 

\textbf{DL software development.} To integrate DL capabilities into software applications, developers make use of state-of-the-art DL frameworks (e.g., TF and Keras) in the software development process. 
Specifically, developers use these frameworks to create the structure of DL models and specify run-time configuration (e.g., hyper-parameters). In a DL model, multiple layers of transformation functions are used to convert input to output, with each layer learning successively higher level of abstractions in the data. Then large-scale data (i.e., the training data) is used to train (i.e., adjust the weights of) the multiple layers. Finally, validation data, which is different from the training data, is used to tune the model. Due to the space limit, we show only the model training phase in Figure~\ref{fig:devordep}.

\textbf{DL software deployment.} After DL software has been well validated and tested, it is ready to be deployed to different platforms for real usage. The deployment process focuses on platform adaptations, i.e., adapting DL software for the deployment platform. The most popular way is to deploy DL software on the server or cloud platforms~\cite{corrabs191003156}. This way enables developers to invoke services powered by DL techniques via simply calling an API endpoint. Some frameworks (e.g., TF Serving~\cite{tfservinglink}) and platforms (e.g., Google Cloud ML Engine~\cite{googlecloudlink}) can facilitate this deployment. In addition, there is a rising demand in deploying DL software to mobile devices~\cite{wwwXuLLLLL19} and browsers~\cite{wwwMaXZTL19}. For mobile platforms, due to their limited computing power, memory size, and energy capacity, models that are trained on PC platforms and used in the DL software cannot be deployed directly to the mobile platforms in some cases. Therefore, some lightweight DL frameworks, such as TF Lite for Android and Core ML for iOS, are specifically designed for converting pre-trained DL models to the formats supported by mobile platforms. In addition, it is a common practice to perform model quantization before deploying DL models to mobile devices, in order to reduce memory cost and computing overhead~\cite{wwwXuLLLLL19,GuoCXMHLLZL19}. For model quantization, TF Lite supports only converting model weights from floating points to 8-bit integers, while Core ML allows flexible quantization modes, such as 32 bits to 16/8/4 bits~\cite{GuoCXMHLLZL19}. For browsers, some solutions (e.g., TF.js~\cite{tfjslink}) are proposed for deploying DL models under Web environments.

\textbf{Scope.} We focus our analysis on DL software deployment. Specifically, we analyze the challenges in deploying DL software to different platforms including server/cloud, mobile, and browser platforms. Any issues related to this process are within our scope. However, challenges related to DL software development (e.g., model training) are not considered in this study.
\section{Methodology}\label{method}
\begin{figure}[t]
\includegraphics[width=1.0\columnwidth]{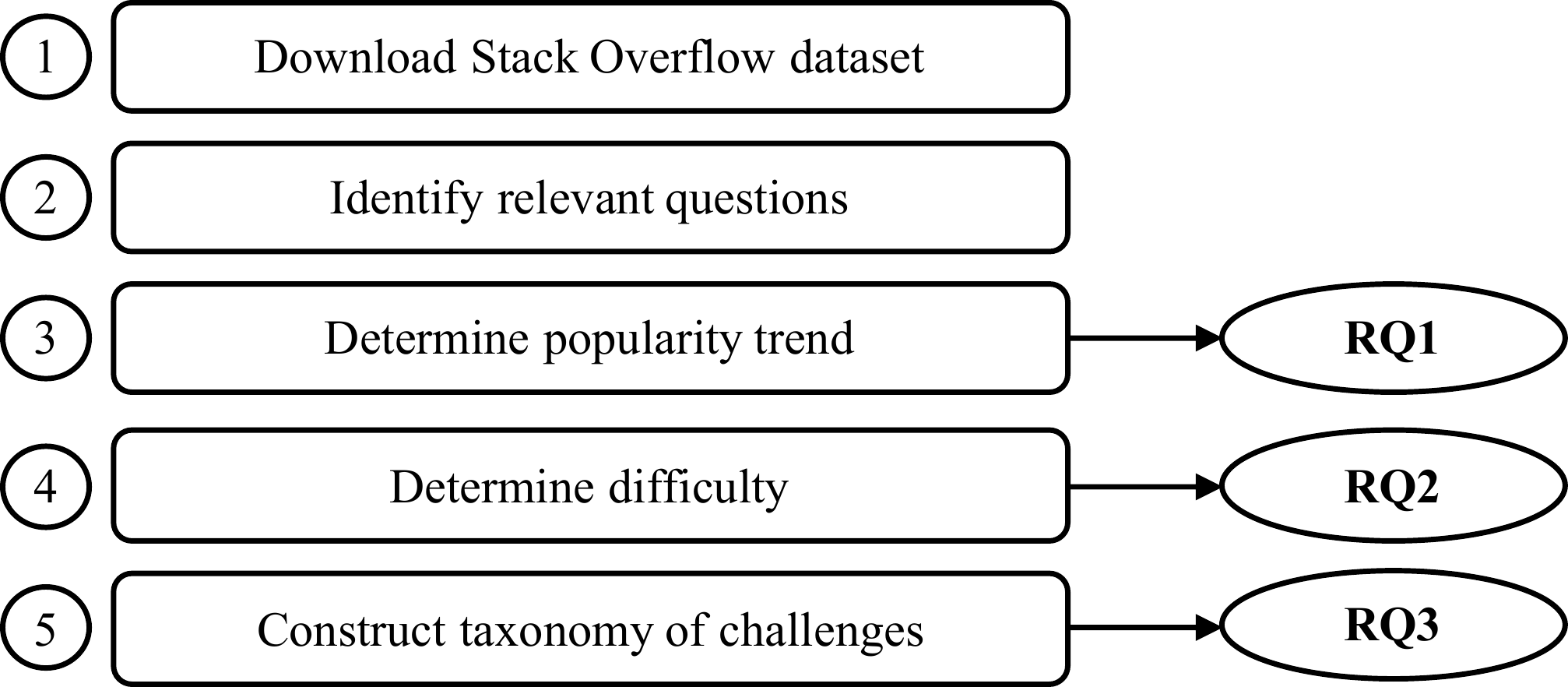}
\caption{An overview of the methodology.}\label{fig:method}
\end{figure}

To understand the challenges in deploying DL software, we analyze the relevant questions posted on Stack Overflow (SO), where developers seek technological advice about unresolved issues. 
We show an overview of the methodology of our study in Figure~\ref{fig:method}.

\para{\emph{Step 1: Download Stack Overflow dataset}} In the first step of this study, we download SO dataset from the official Stack Exchange Data Dump~\cite{sodump} on December 2, 2019. The dataset covers the SO posts generated from July 31, 2008 to December 1, 2019.
The meta data of each post includes its identifier, post type (i.e., question or answer), creation date, tags, title, body, identifier of the accepted answer if the post is a question, etc. Each question has one to five tags based on its topics. The developer who posted a question can mark an answer as an accepted answer to indicate that it works for the question. Among all the questions in the dataset (denoted as the set $\mathcal{A}$), 52.33\% have an accepted answer.

\para{\emph{Step 2: Identify relevant questions}}
In this study, we select three representative deployment platforms of DL software for study, including server/cloud, mobile, and browser platforms. Since questions related to DL software deployment may be contained in DL related questions, we first identify SO questions related to DL. Following previous work~\cite{sigsoftIslamNPR19,corrabstaxonomy}, we extract questions tagged with at least one of the top five popular DL frameworks (i.e., TF, Keras, PyTorch, Theano, and Caffe) from $\mathcal{A}$ and denote the extracted 70,669 questions as the set $\mathcal{B}$. Then we identify the relevant questions for each kind of platform, respectively.


\textbf{Server/Cloud.} We first define a vocabulary of words related to server/cloud platforms (i.e., ``cloud'', ``server'', and ``serving''). Then we perform a case-insensitive search of the three terms within the title and body (excluding code snippets) of each question in $\mathcal{B}$ and denote the questions that contain at least one of the terms as the set $\mathcal{C}$. Since questions in $\mathcal{C}$ may contain some noise that is not related to deployment (e.g., questions about training DL models on the server), we filter out those that do not contain the word ``deploy'' and finally 279 questions remain in $\mathcal{C}$. To further complement $\mathcal{C}$, we extract questions tagged with TF Serving, Google Cloud ML Engine, and Amazon SageMaker from $\mathcal{A}$. TF Serving is a DL framework that is specifically designed for deploying DL software to servers; Google Cloud ML Engine and Amazon SageMaker~\cite{amazonlink} are two popular cloud platforms for training DL models and deploying DL software. Since the two platforms are rolled out by two major cloud service vendors, i.e., Google and Amazon, we believe that they are representative. For questions tagged with the two platforms, we filter out those that do not contain the word ``deploy'' as they also support model training.
Then we add the remaining questions as well as all questions tagged with TF Serving into $\mathcal{C}$ and remove the duplicate questions. Finally, we have 1,325 questions about DL software deployment to server/cloud platforms in the set $\mathcal{C}$.

\textbf{Mobile.} We define a vocabulary of words related to mobile devices (i.e., ``mobile'', ``android'', and ``ios'') and extract the questions that contain at least one of the three words from $\mathcal{B}$ in a case-insensitive way. We denote the extracted 486 questions as the question set $\mathcal{D}$. Then, following previous work~\cite{GuoCXMHLLZL19}, we also consider two DL frameworks specifically designed for DL software deployment to mobile platforms (i.e., TF Lite and Core ML). We extract the questions tagged with at least one of the two frameworks from $\mathcal{A}$ and then add these questions into $\mathcal{D}$. Finally, we remove the duplicate questions and have 1,533 questions about DL software deployment to mobile platforms in the set $\mathcal{D}$.

\textbf{Browser.} We extract questions that contain the word ``browser'' from $\mathcal{B}$ in a case-insensitive way and denote the extracted 89 questions as the set $\mathcal{E}$. In addition, following previous work~\cite{GuoCXMHLLZL19}, we also take TF.js, which can be used for deploying DL models on browsers, into consideration. Different from TF Lite, which supports only deployment, TF.js also supports developing DL models. However, since DL on browsers is still at dawn~\cite{wwwMaXZTL19}, questions tagged with TF.js in $\mathcal{A}$ are too few, only 535. If we employ strict keyword matching  to filter out questions that do not contain ``deploy'' as above, only 10 out of 535 questions can remain. To keep as many relevant questions as possible, instead of keyword matching, we employ manual inspection here. Specifically, we add all the 535 questions into $\mathcal{E}$ and exclude the duplicate questions. Then two authors of this paper examine the remaining 576 questions independently and determine whether or not each question is about DL software deployment. The inter-rater agreement measured as Cohen's Kappa ($\kappa$)~\cite{cohen1960coefficient} is 0.894, which indicates almost perfect agreement. Then the conflicts are resolved through discussion, and the questions considered as non-deployment issues are excluded from $\mathcal{E}$. Finally, we have 165 questions about DL software deployment to browser 
platforms in the set $\mathcal{E}$.

\para{\emph{Step 3: Determine popularity trend}}
To illustrate the popularity trend of DL software deployment, following previous work~\cite{AlshangitiSMLY19}, we calculate the number of users and questions related to the topic per year. 
Specifically, the metrics are calculated based on the question sets $\mathcal{C}$, $\mathcal{D}$, and $\mathcal{E}$,  for each of the past five years (i.e., from 2015 to 2019). Step 3 answers the research question \textbf{RQ1}.

\para{\emph{Step 4: Determine difficulty}}
We measure the difficulty of deploying DL software using two metrics widely adopted by previous work~\cite{AlshangitiSMLY19,sigsoftBagherzadehK19,esemAhmedB18}, including the percentage of questions with no accepted answer (``\emph{\%no acc.}'') and the response time needed to receive an accepted answer. In this step, we use the questions related to other aspects of DL software (in short as \emph{non-deployment questions}) as the baseline for comparison. To this end, we exclude the deployment related questions (i.e., questions in $\mathcal{C}$, $\mathcal{D}$, and $\mathcal{E}$) from the DL related questions (i.e., questions in $\mathcal{B}$), and use the remaining questions as the non-deployment questions. For the first metric, we employ proportion test~\cite{NewcombeInterval} to ensure the statistical significance of comparison. This test is used for testing null hypothesis that proportions in multiple groups are the same~\cite{portest}, and thus the test is appropriate for the comparison in \emph{\%no acc.}
For the second metric, we select the questions that have received accepted answers and then show the distribution and the median value of the response time needed to receive an accept answer for both deployment and non-deployment questions. Step 4 answers the research question \textbf{RQ2}.

\para{\emph{Step 5: Construct taxonomy of challenges}} In this step, we manually analyze the questions related to DL software deployment, in order to construct the taxonomy of challenges.
Following previous work~\cite{issredeep19}, to ensure a 95\% confidence level and a 5\% confidence interval, we randomly sample 297 server/cloud related questions from $\mathcal{C}$ and 307 mobile related questions from $\mathcal{D}$. Since browser related questions in $\mathcal{E}$ are not too many, we use all the 165 questions in it for manual analysis. In total, we get a dataset of 769 questions that are used for taxonomy construction. The size of this dataset is comparable and even larger than those used in existing studies~\cite{isstaZhangCCXZ18,issredeep19,iwpcBeyerM0P18,icseAghajaniNVLMBL19} that also require manual analysis of SO posts. 
Next, we present our procedures of taxonomy construction.

\textbf{Pilot construction.} First, we randomly sample 30\% of the 769 questions for a pilot construction of the taxonomy. The taxonomy for each kind of platform is constructed individually based on its corresponding samples. We follow an open coding procedure~\cite{tseSeaman99} to inductively create the categories and subcategories of our taxonomy in a bottom-up way by analyzing the sampled questions. The first two authors (named as inspectors), who both have four years of DL experiences, jointly participate in the pilot construction. The detailed procedure is described below.

The inspectors read and reread all the questions, in order to be familiar with them. In this process, the inspectors take all the elements of each question, including the title, body, code snippets, comments, answers, tags, and even URLs mentioned by questioners and answerers, for careful inspection. Questions not related to DL software deployment are classified as \emph{False positives}. For a relevant question, if the inspectors cannot identify the specific challenge behind it, they mark it as \emph{Unclear questions}, which as well as \emph{False positives} are not included into the taxonomy. For the remaining questions, the inspectors assign short phrases as initial codes to indicate the challenges behind these questions. Specifically, for those questions that are raised without attempts (mainly in the form of ``how'', e.g., ``\textit{how to process raw data in tf-serving}''~\cite{so8}), the inspectors can often clearly identify the challenges from the question descriptions; for those questions that describe the faults or unexpected results that developers encounter in practice, the inspectors identify their causes as the challenges. For example, if a developer reports an error that she encounters when making predictions and the inspectors can find that the cause is the wrong format of input data from the question descriptions, comments, or answers, the inspectors consider setting the format of input data as the challenge behind this question.

Then the inspectors proceed to group similar codes into categories and create a hierarchical taxonomy of challenges. The grouping process is iterative, in which the inspectors continuously go back and forth between categories and questions to refine the taxonomy. A question is assigned to all related categories if it is related to multiple challenges. All conflicts are discussed and resolved by introducing three arbitrators. The arbitrator for server/cloud deployment is a practitioner who has four years of experience in deploying DL software to servers/cloud platforms. The two arbitrators for mobile and browser deployment are graduate students who have two years of experience in deploying DL software to mobile devices and browsers, respectively. Both of these arbitrators have published papers related to DL software deployment in top-tier conferences. The arbitrators finally approve all categories in the taxonomy.

 \textbf{Reliability analysis and extended construction.} Based on the coding schema in the pilot construction, the first two authors then independently label the remaining 70\% questions for reliability analysis. Each question is labeled with \emph{False positives}, \emph{Unclear questions}, or the identified leaf categories in the taxonomy. Questions that cannot be classified into the current taxonomy are added into a new category named \emph{Pending}. The inter-rater agreement during the independent labeling is 0.816 measured by Cohen's Kappa ($\kappa$), indicating almost perfect agreement and demonstrating the reliability of our coding schema and procedure. The conflicts of labeling are then discussed and resolved by the aforementioned three arbitrators. For the questions classified as \emph{Pending}, the arbitrators help further identify the challenges behind the questions and determine whether new categories need to be added. Finally, 8 new leaf categories are added and all questions in \emph{Pending} are assigned into the taxonomy. 
 
In summary, among the 769 sampled questions, 58 are marked as \emph{False positives}, and 130 as  \emph{Unclear questions}. In addition, there are 2 questions each of which is assigned into two categories. The remaining 583 sampled questions (i.e., 227 for server/cloud deployment, 231 for mobile deployment, and 125 for browser deployment) are all covered in the final taxonomy.  The entire manual construction process takes about 450 man-hours. Step 5 answers the research question \textbf{RQ3}.


\section{RQ1: Popularity Trend}
\begin{figure}[!tp]
    \centering
    \subfigure[Trend of users]{
        \label{fig:user_trend}
        \includegraphics[width=0.2\textwidth]{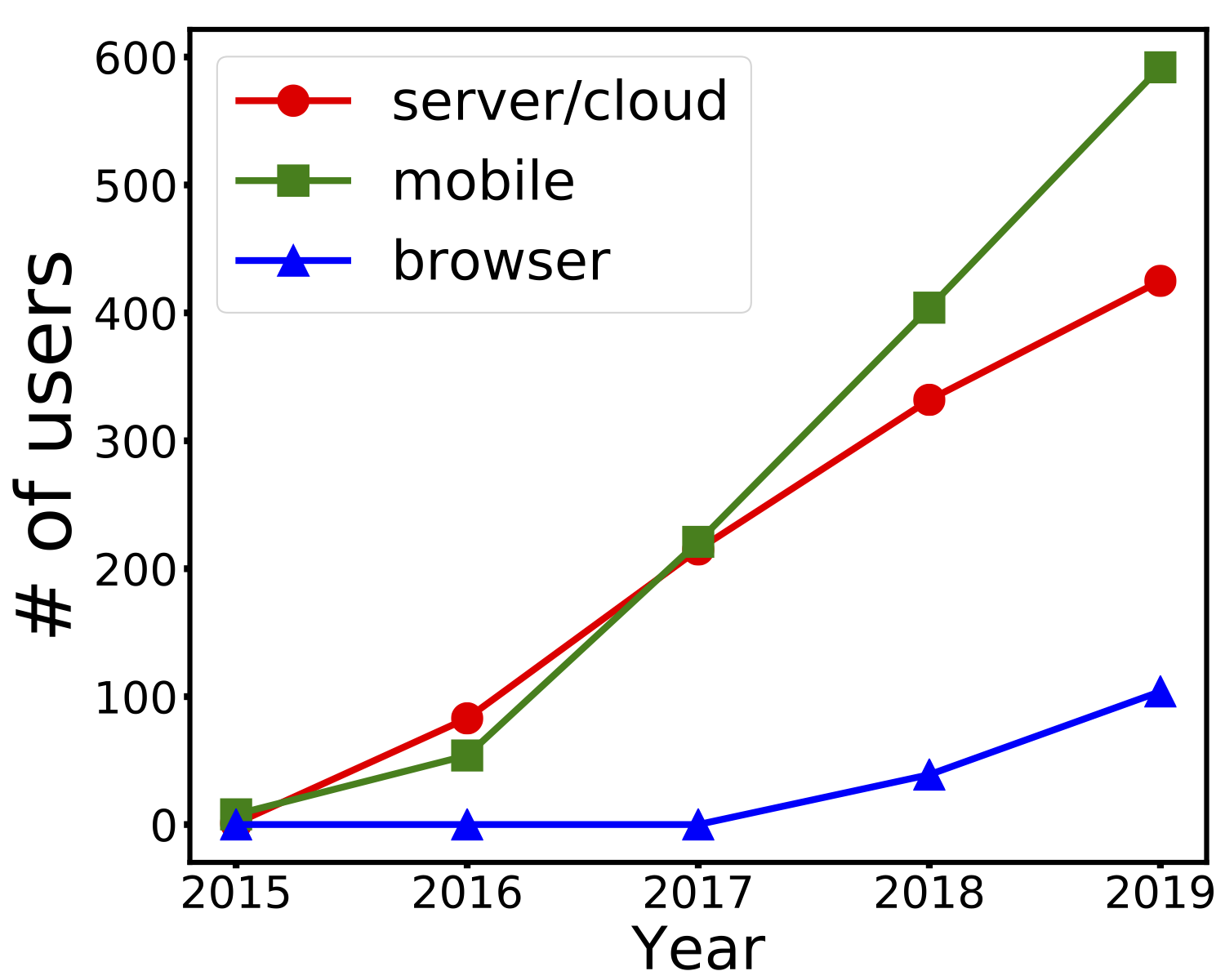}
    }
    \subfigure[Trend of questions]{
        \label{fig:question_trend}
        \includegraphics[width=0.2\textwidth]{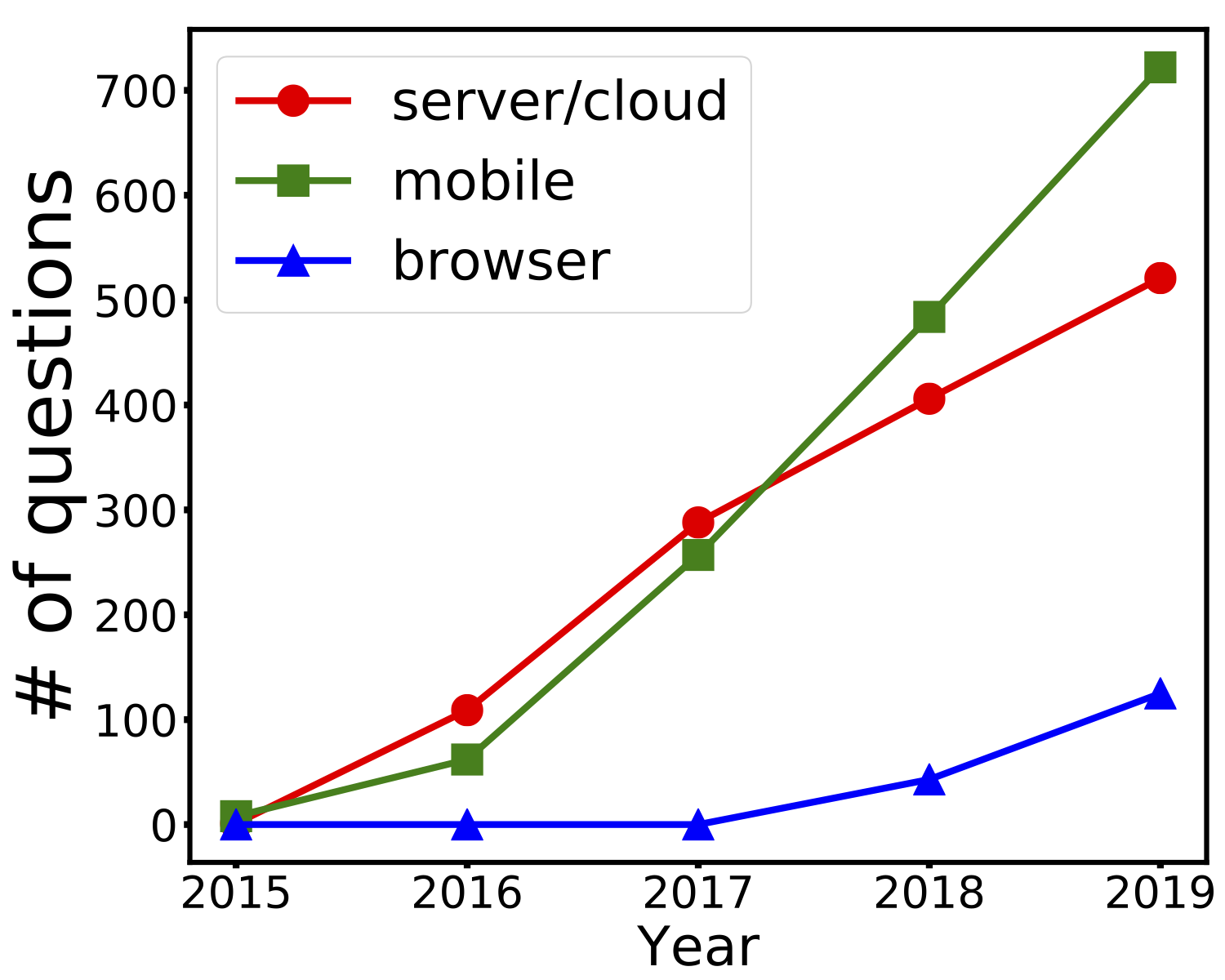}
    }
    \caption{The popularity trend of deploying DL software.}\label{fig:dl_trend}
\end{figure}

Figure~\ref{fig:dl_trend} shows the popularity trend of deploying DL software in terms of the number of users and questions on SO. 
The figure indicates that this topic is gaining increasing attention, demonstrating the timeliness and urgency of this study.


For deploying DL software on server/cloud platforms, we observe that users and questions increase in a steady trend. In 2017, most major vendors roll out their DL frameworks for mobile devices~\cite{wwwXuLLLLL19}. As a result, we can observe that both the number of users and the number of questions related to mobile deployment in 2017 increase by more than 300\% compared to 2016. 
For deploying DL software on browsers, questions start to appear in 2018 due to the release of TF.js in 2018. As found by Ma et al.~\cite{wwwMaXZTL19}, DL in browsers is still at dawn. Therefore, the users and questions related to DL are still not so many, as shown in Figure~\ref{fig:dl_trend}.

\section{RQ2: Difficulty}
For deployment and other aspects (in short of \emph{non-deployment}) of DL software, the percentages of relevant questions with no accepted answer (\emph{\%no acc.}) are 70.7\% and 62.7\%, respectively.
The significance of this difference is ensured by the result of proportion test ($\chi^2 = 78.153$, df = 1, $p$-value $\textless$ 2.2e-16), indicating that questions related to DL software deployment are more difficult to answer than those related to other aspects of DL software. More specifically, for server/cloud, mobile, and browser deployment, the values of \emph{\%no acc.} are 69.8\%, 71.6\%, and 69.1\%, respectively. In terms of this metric, questions about deploying DL software are also more difficult to resolve than other well-studied challenging topics in SE, such as big data (\emph{\%no acc.} = 60.5\%~\cite{sigsoftBagherzadehK19}), concurrency (\emph{\%no acc.} = 43.8\%~\cite{esemAhmedB18}), and mobile (\emph{\%no acc.} = 55.0\%~\cite{eseRosenS16}).

\begin{figure}
\includegraphics[width=0.6\columnwidth]{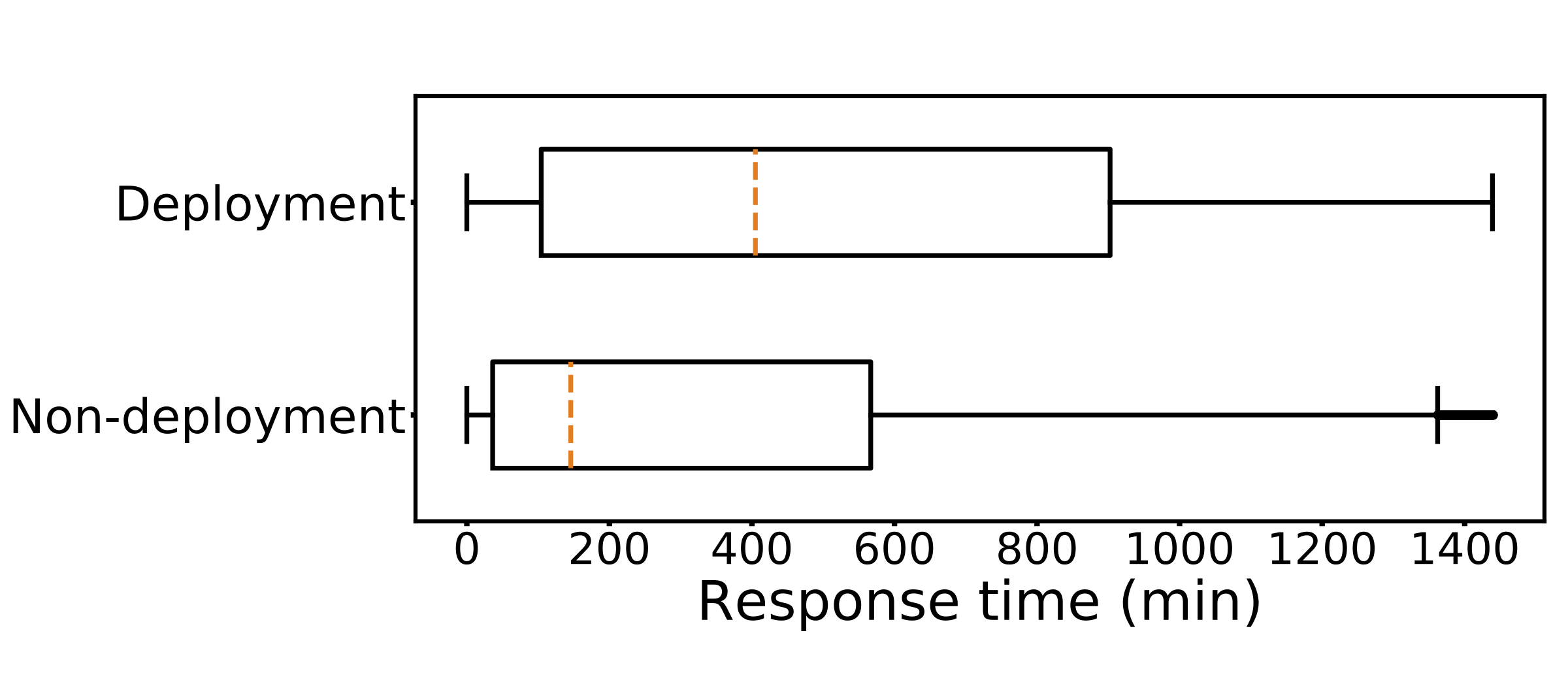}
\caption{Time needed to receive an accepted answer.}\label{fig:time_acc}
\end{figure}

Figure~\ref{fig:time_acc} presents the boxplot of response time needed to receive an accepted answer for deployment and non-deployment related questions. We can observe that the time needed for non-deployment questions is mostly concentrated below 600 minutes, while deployment questions have a wider spread.
Furthermore, we find that the median response time for deployment questions (i.e., 404.9 minutes) is about 3 times the time needed for non-deployment questions (only 145.8 minutes). More specifically, the median response time for server/cloud, mobile, and browser related questions is 428.5, 405.2, and 180.9 minutes, respectively. In previous work~\cite{sigsoftBagherzadehK19,esemAhmedB18,eseRosenS16}, researchers find that the median response time needed for other challenging topics, including big data, concurrency, and mobile, is about 198~\cite{sigsoftBagherzadehK19}, 42~\cite{esemAhmedB18}, and 55 minutes~\cite{eseRosenS16}, respectively. In contrast, questions related to deploying DL software need longer time to receive accepted answers.

In summary, we find that questions related to DL software deployment are difficult to resolve, partly demonstrating the finding in previous work~\cite{AlshangitiSMLY19} that model deployment is the most challenging phase in the life cycle of machine learning (ML) and motivating us to further identify the specific challenges behind deploying DL software.

\section{RQ3: Taxonomy of Challenges}
\begin{figure*}
\includegraphics[width=2\columnwidth]{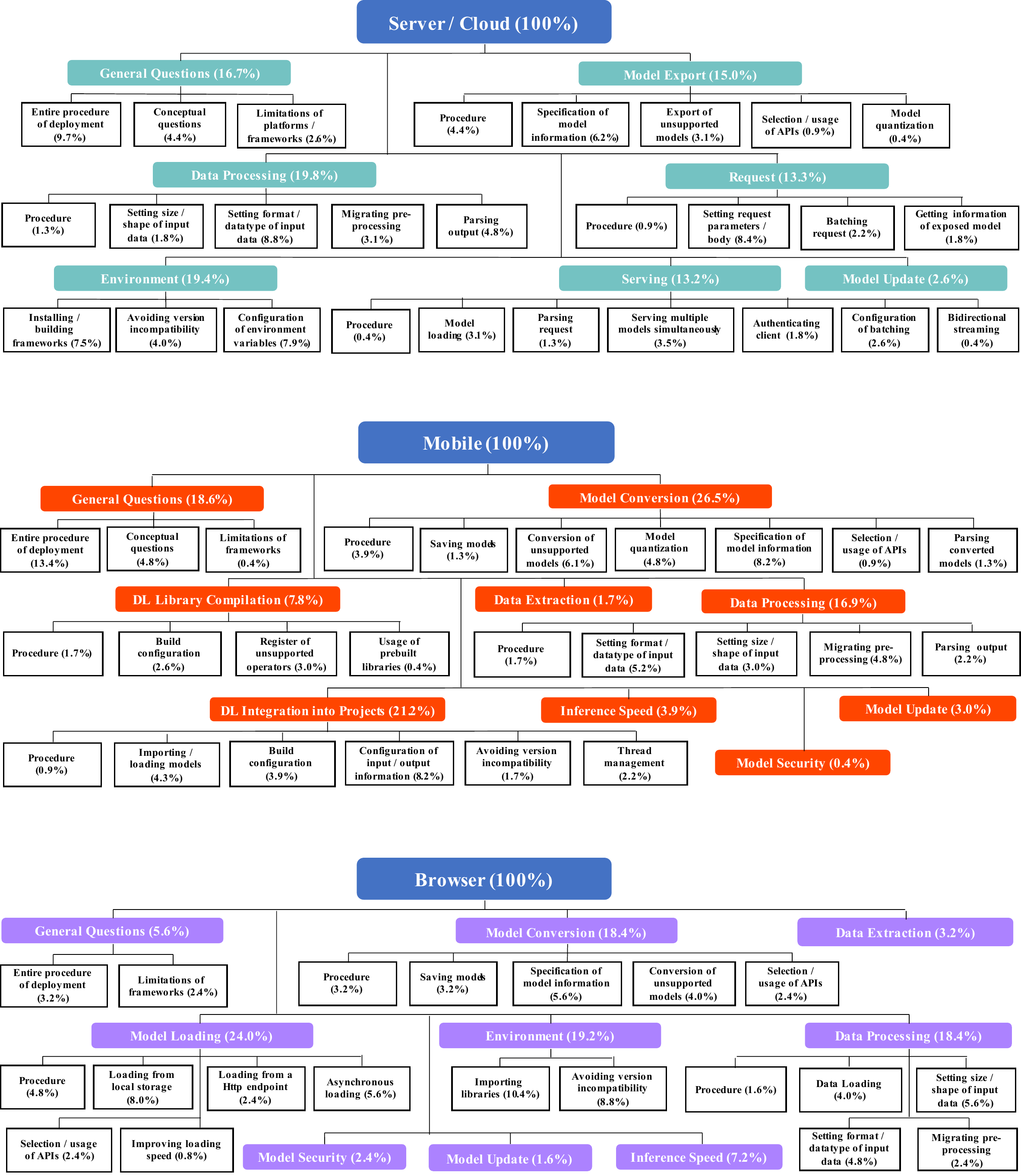}
\caption{Taxonomy of challenges in deploying DL software.}\label{fig:tax}
\end{figure*}

Figure~\ref{fig:tax} illustrates the hierarchical taxonomy of challenges in DL software deployment. As shown in Figure~\ref{fig:tax}, developers have difficulty in a broad spectrum of issues. Note that although the identified challenges are about deploying DL software to specific platforms, not all relevant issues occur on corresponding platforms. For example, to deploy DL software to mobile devices, the model conversion task can be done on PC platforms. 

We group the full taxonomy into three sub-taxonomies that correspond to the challenges in deploying DL software to server/cloud, mobile, and browser platforms, respectively. Each sub-taxonomy is then organized into three-level categories, including the root categories (e.g., \emph{Server/Cloud}), the inner categories (e.g., \emph{Model Export}), and the leaf categories (e.g., \emph{Model quantization}). In total, we have 3 root categories, 25 inner categories, and 72 leaf categories. We show the percentages for questions related to each category in the parentheses. Then we describe and exemplify each inner category.

\subsection{Common Challenges in Server/Cloud, Mobile, and Browser}

To avoid duplicate descriptions, we first present the common inner categories in \emph{Server/Cloud}, \emph{Mobile}, and \emph{Browser}.

\subsubsection{General Questions.} This category shows general challenges that do not involve a specific step in the deployment process, and contains several leaf categories as follows.

\textbf{Entire procedure of deployment.} This category refers to general questions about the entire procedure of deployment,  mainly raised without practical attempts. These questions are mainly in the form of ``how'', such as ``\textit{how can I use that model in android for image classification}''~\cite{so1}.
In such questions, developers often complain about the documentation, e.g., ``\textit{there is no documentation given for this model}''~\cite{so62}.
Answerers mainly handle these questions by providing existing tutorials or documentation-like information that does not appear elsewhere, or translate the jargon-heavy documentation into case-specific guidance phrased in a developer-friendly way.
Compared to \emph{Server/Cloud} (9.7\%) and \emph{Mobile} (13.4\%), \emph{Browser} contains relatively fewer such questions (3.2\%). A possible explanation is that since DL in browsers is still in the early stage~\cite{wwwMaXZTL19}, developers are mainly stuck in DL's primary usage rather than being eager to explore how to apply DL to various scenarios. 

\textbf{Conceptual questions.} This category includes questions about basic concepts or background knowledge related to DL software deployment, such as ``\textit{is there any difference between these Neural Network Classifier and Neural Network in machine learning model type used in iOS }''~\cite{so3}. This category of questions is also observed in previous work~\cite{esemAhmedB18,sigsoftBagherzadehK19,chiprivacy} that analyzes challenges faced by developers in other topics through SO questions.
For \emph{Server/Cloud} and \emph{Mobile}, this category accounts for 4.4\% and 4.8\%, respectively, indicating that developers find even the basics of DL software deployment challenging. 
For \emph{Browser}, this category is missing. Since TF.js also supports model training, we filter out the conceptual questions about TF.js during manual inspection as we cannot discern whether these questions occur during training or deployment. However, it does not mean that there is no conceptual questions about browser deployment. 

\textbf{Limitations of platforms/frameworks.} This category is about limitations of relevant platforms or DL frameworks. For example, a senior software engineer working on the Google Cloud ML Platform team apologizes for the failure that a developer encounters, admitting that the platform currently does not support batch prediction~\cite{so4}. Besides, some issues reflect bugs in current deployment related frameworks. For instance, an issue reveals a bug in the \textit{TocoConvert.from\_keras\_model\_file} method of TF Lite~\cite{so5}. 

\subsubsection{Model Export and Model Conversion} Both categories cover challenges in converting DL models in DL software into the formats supported by deployment platforms. Model export directly saves the trained model into the expected format, and it is a common way for deploying DL models to server/cloud platforms. In contrast, model conversion  needs two steps: (1) saving the trained model into a format supported by the deployment frameworks; (2) using these frameworks to convert the saved model into the format supported by mobile devices or browsers. Considering the similar functionalities of model export and model conversion, we put them together for description. Model export represents 15.0\% of questions in \emph{Server/Cloud}, while model conversion covers the most encountered category of challenges in \emph{Mobile} and the third most encountered category of challenges in \emph{Browser}, accounting for 26.5\% and 18.4\%, respectively. We next present representative leaf categories under the two categories.

\textbf{Procedure.} Different from \emph{Entire procedure of deployment}, which asks about the entire deployment process, questions in \emph{Procedure} are about the procedure of a specific step in the process. An example question in \emph{Procedure} under \emph{Model Conversion} is ``\textit{how can I convert this file into a .coreml file}''~\cite{so61}. Due to page limit, we do not repeat the descriptions of \emph{Procedure} in other inner categories.

\textbf{Export/conversion of unsupported models.} The support of DL on some platforms is still unfledged. Some standard operators and layers used in the trained model are not supported by deployment frameworks. For example, developers report that \emph{LSTM} is not supported by TF Lite~\cite{so23} and that \emph{GaussianNoise} is not supported by TF.js~\cite{so24}. Similarly,  Guo et al.~\cite{GuoCXMHLLZL19} report that they could not deploy the RNN models (i.e., LSTM and GRU) to mobile platforms due to the ``\emph{unsupported operation}'' error. In addition, when developers attempt to export or convert models with custom operators or layers, the developers also encounter difficulties~\cite{so25,so26}.

\textbf{Specification of model information.} When exporting or converting DL models to expected formats, developers  need to specify model information. For instance, TF Serving requires developers to construct a signature to specify names of the input and output tensors and the method of inference (i.e., regression, prediction, or classification)~\cite{do2}. Incorrect specification would result in errors~\cite{so27}. Sometimes, developers directly use off-the-shelf models that have been well trained and released online for deployment, but the developers have no idea about the models' information (e.g., names of the input and output tensors~\cite{so28}), making the model export/conversion task challenging.

\textbf{Selection/usage of APIs.} There are so many APIs provided by different frameworks for developers to export and convert models to various formats. Therefore, it is challenging for developers to select and use these APIs correctly according to their demand. For example, a developer is confused about the ``\textit{relationship between tensorflow saver, exporter and save model}''~\cite{so29} and says frankly that she feels more confused after reading some tutorials. In addition, the addition, deprecation, and upgrade of APIs caused by the update of frameworks also make the selection and usage of APIs error-prone~\cite{so30}. 

\textbf{Model quantization.} Model quantization reduces precision representations of model weights, in order to reduce memory cost and computing overhead of DL models~\cite{do3}. It is mainly used for deployment to mobile devices, due to their limitations of computing power, memory size, and energy capacity. For this technique, developers have difficulty in configuration of relevant parameters~\cite{so31}. In addition, developers call for support of more quantization options. For instance, TF Lite supports only 8-bit quantization (i.e., converting model weights from floating points to 8-bit integers), but developers may need more bits for quantization~\cite{so32}. 

\subsubsection{Data Processing.} This category covers challenges in converting raw data into the input format needed by DL models in DL software (i.e., pre-processing) and converting the model output into expected formats (i.e., post-processing). This category accounts for the most questions (19.8\%) in \emph{Server/Cloud}. For \emph{Mobile} and \emph{Browser}, it covers 16.9\% and 18.4\% of questions, respectively. We next describe the representative leaf categories under \emph{Data Processing}.

\textbf{Setting size/shape/format/datatype of input data.} It is a common challenge in data pre-processing to set the size/shape and format/datatype of data. A faulty behavior manifests when the input data has an unexpected size/shape (e.g., a \emph{224$\times$224 image} instead of a \emph{227$\times$227 image}~\cite{so10}), format (e.g., encoding an image in the \emph{Base64 format} instead of converting it to \emph{a list} ~\cite{so11}), or datatype (e.g., \emph{float} instead of \emph{int}~\cite{so12}). 

\textbf{Migrating pre-processing.} When  ML/DL models are developed, data pre-processing is often considered as an individual phase~\cite{AlshangitiSMLY19} and thus may not be included inside the model structure. In this case, code for data pre-processing needs to be migrated during the deployment process, in order to keep consistent behaviors of software before and after deployment. For instance, when developers deploy a DL application with pre-processing implemented with Python and out of the DL model to an Android device, the developers may need to re-implement pre-processing using a new language (e.g., Java or C/C++). Forgetting to re-implement pre-processing~\cite{so13} or re-implementing it incorrectly~\cite{so15} can lead to faulty behaviors. In addition, an alternative to keep data pre-processing consistent is to add it into the structure of DL models. For this option, developers face challenges such as ``\textit{how to add layers before the input layer of model restored from a .pb file [...] to decode jpeg encoded strings and feed the result into the current input tensor}''~\cite{so14}.

\textbf{Parsing output.} This category includes challenges in converting the output of DL models to expected or human-readable results, such as parsing the output array~\cite{so17} or tensor~\cite{so16} to get the actual predicted class.

\subsubsection{Model Update.} Once DL software is deployed for real usage, the DL software can receive feedback (e.g., bad cases) from users. The feedback can be used to update weights of the DL model in DL software for further performance improvement.
Many challenges, such as periodical automated model update on clouds~\cite{so18} and model update (or re-training) on mobile devices~\cite{so19}, emerge from the efforts to achieve this goal.
This category covers 2.6\%, 3.0\%, and 1.6\% of questions in \emph{Server/Cloud}, \emph{Mobile}, and \emph{Browser}, respectively.

\subsection{Common Challenges in Mobile and Browser} 

\subsubsection{Data Extraction.} 
To deploy DL software successfully, developers need to consider any stage that may affect the final performance, including data extraction.
This category is observed only in \emph{Mobile} and \emph{Browser}, accounting for 1.7\% and 3.2\% of questions, respectively. This finding indicates the difficulty of extracting data in mobile devices and browsers.

\subsubsection{Inference Speed.} Compared to server/cloud platforms, mobile and browser platforms have weaker computing power. As a result, the inference speed of the deployed software has been a challenge in mobile devices (3.9\%) and browsers (7.2\%). 

\subsubsection{Model Security.} DL models in DL software are often stored in unencrypted formats, resulting in a risk that competitors may disassemble and reuse the models. To alleviate this risk and ensure model security, developers attempt multiple approaches, such as obfuscating code~\cite{so6} or libraries~\cite{so7}. Any challenges related to model security are included in this category. This category is observed only in \emph{Mobile} and \emph{Browser}, since models deployed to these platforms are easier to obtain. In contrast, models deployed on server/cloud platforms are hidden behind API calls.

\subsection{Common Challenges in Server/Cloud and Browser}\label{inner_sb}
\emph{Environment.} This category includes challenges in setting up the environment for DL software deployment, and accounts for 19.4\% and 19.2\% of questions in \emph{Server/Cloud} and \emph{Browser}, respectively. For \emph{Mobile}, its environment related questions are mainly distributed in \emph{DL Library Compilation} and \emph{DL Integration into Projects} categories that will be introduced later. 
When deploying DL software to server/cloud platforms, developers need to configure various environment variables, whose diverse options make the configuration task challenging. In addition, for the server deployment, developers also need to install or build necessary frameworks such as TF Serving. Issues that occur in this phase are included in \emph{Installing/building frameworks}. Similarly, when deploying DL software to browsers, some developers have difficulty in \emph{Importing libraries}, e.g., ``\textit{I am developing a chrome extension, where I use my trained keras model. For this I need to import a library tensorflow.js. How should I do that}''~\cite{so33}. Besides these challenges, the rapid evolution of DL frameworks makes the version compatibility of frameworks/libraries challenging for developers. For instance, an error reported on SO is caused by that the TF used to train and save the model has an incompatible version with TF Serving used for deployment~\cite{so35}. Similarly, Humbatova et al.~\cite{corrabstaxonomy} mention that version incompatibility between different libraries and frameworks is one of the main concerns of practitioners in developing DL software.

\subsection{Remaining Challenges in Server/Cloud}

\subsubsection{Request.} This category covers challenges in making requests in the client and accounts for 13.3\% of questions in \emph{Server/Cloud}. For \emph{Request}, developers have difficulty in configuring the request body~\cite{so47}, sending multiple requests at a single time (i.e., batching request)~\cite{so48}, getting information of serving models via request~\cite{so46}, etc.
 
\subsubsection{Serving.} This category concerns challenges related to serving DL software on the server/cloud platforms and accounts for 13.2\% of questions. 
To make a DL model in DL software servable, developers first need to load the DL model, where issues such as loading time~\cite{so49} and memory usage~\cite{so50} may emerge. In addition, many developers encounter difficulties in authenticating the client~\cite{so51} and parsing the request~\cite{so52}. Sometimes, developers need to serve multiple different models to provide diverse services or serve different versions of one model at the same time~\cite{so53}, but the developers find that the implementation is not such easy (accounting for 3.5\% of questions). Similarly, Zhang et al.~\cite{corrabs191003156} demonstrate that multiple-model maintenance is one of the main challenges in DL software deployment and maintenance in the server side. Finally, we want to mention a specific configuration issue in this category, i.e., \emph{Configuration of batching}. To process requests in batches, developers need to configure relevant parameters manually. We observe this issue in 2.6\% of questions, e.g., ``\textit{I know that the batch.config file needs to be fine-tuned a bunch by hand, and I have messed with it a lot and tuned numbers around, but nothing seems to actually effect runtimes}''~\cite{so36}.

\subsection{Remaining Challenges in Mobile}
\subsubsection{DL Library Compilation.} This category includes challenges in compiling DL libraries for target mobile devices and covers 7.8\% of questions in \emph{Mobile}. Since Core ML is well supported by iOS, developers can use Core ML directly without installing or building it. For TF Lite, pre-built libraries are officially provided for developers’ convenience. However, developers still need to compile TF Lite from source code by themselves in some cases (e.g., deploying models containing unsupported operators). Since the operators supported by TF Lite are still insufficient to meet developers’ demand~\cite{so38}, developers sometimes need to register unsupported operators manually to add them into the run-time library. It may be challenging for developers who are unfamiliar with TF Lite. In addition, for compilation, developers need to configure build command lines and edit configuration files (i.e., \emph{Build configuration}). Wrong configurations~\cite{so39} can result in build failure or library incompatibility with target platforms.

\subsubsection{DL Integration into Projects.} This category includes challenges in integrating DL libraries and models into mobile software projects. It accounts for 21.2\% in \emph{Mobile}. 
To integrate DL libraries and build projects, developers need to edit build configuration files (i.e., \emph{Build configuration}), being a common challenge (3.9\%) for both Android and iOS developers. 
To integrate DL models into projects, developers face challenges in importing and loading models (4.3\%). For example, in an Xcode project for iOS, developers can drag a model into the project navigator, and then Xcode can parse and import the model automatically~\cite{so41}. However, some developers encounter errors during this process~\cite{so54,so55}. When it comes to an Android project, the importing process is more complicated. For instance, if developers load a TF Lite model with C++ or Java, they need to set the information (e.g., datatype and size) of input and output tensors manually (8.2\%), but some developers fail in this configuration~\cite{so44}. In addition, developers have difficulty in the thread management (2.2\%) when integrating DL models into projects, e.g., ``\textit{I am building an Android application that has three threads running three different models, would it be possible to still enable inter\_op\_parallelism\_threads and set to 4 for a quad-core device}''~\cite{so56}.

\subsection{Remaining Challenges in Browser}\label{un_brow}

\emph{Model Loading.} This category includes challenges in loading DL models in browsers, being the most common challenges in browser deployment (accounting for 24.0\% of questions). For browsers, TF.js provides a \emph{tf.loadLayersModel} method to support loading models from local storage, Http endpoints, and IndexedDB. Among the three ways, we observe that the main challenge lies in loading from local storage (8.0\%). In the official document of TF.js~\cite{do4}, ``\emph{local storage}'' refers to the browser's local storage, which is interpreted in a hyperlink~\cite{do6} contained in the document as that ``\emph{the stored data is saved across browser sessions}.'' However, nearly all bad cases in \emph{Loading from local storage} attempt to load models from local file systems. In fact, \emph{tf.loadLayersModel} uses the \emph{fetch} method~\cite{do5} under the hood. \emph{Fetch} is used to get a file served by a server and cannot be used directly with local files. To work with local files, developers first need to serve them on a server.
In addition, many developers do not have a good grasp of the asynchronous loading (5.6\%). In a scenario, when a developer loads a DL model in Chrome and then uses the model to make predictions, she receives  ``\textit{loadedModel.predict is not a function error}'' since the model has not been successfully loaded~\cite{so37}. Since model loading is an asynchronous process in TF.js, developers need to either use \emph{await} or \emph{.then} to wait for the model to be completely loaded before using it for further actions.

\subsection{Unclear Questions}\label{unclear}
Although unclear questions are not included in our taxonomy, we also manually examine them to seek for some insights. All unclear questions have no accepted answers and do not have informative discussions or question descriptions to help us determine the challenges behind the questions. Among these unclear questions, 53\% report unexpected results~\cite{so22} or errors~\cite{so21} when making predictions using the deployed models. However, no anomalies occur at any phase before the phase of making predictions, making it rather difficult to discover the underlying challenges. In fact, various issues can result in the errors or unexpected results in this phase. Take the server deployment as an example. During the manual inspection, we find that errors occurring in making predictions can be attributed to the improper handling of various challenges, such as version incompatibility between libraries used for training and deployment~\cite{so58} (i.e.,  \emph{Environment}), wrong specification of model information~\cite{so59} (i.e., \emph{Model Export}), mismatched format of input data~\cite{so60} (i.e., \emph{Data Processing}), etc.


\section{Implications}\label{imply}
Based on the preceding derived findings, we next discuss our insights and some practical implications for developers, researchers, and DL framework vendors.

\subsection{Researchers}
As demonstrated in our study, DL software deployment is gaining increasing attention from developers, but developers encounter a spectrum of challenges and various unresolved issues. These findings encourage researchers to develop technology to help developers meet these deployment challenges. Here, we briefly discuss some potential opportunities to the research communities based on our results.
\textbf{(1) Automated fault localization.} In Section~\ref{unclear}, we find that 53\% of unclear questions report errors when making predictions and that various faults in different phases can result in such errors. This finding indicates the difficulty in manually locating faults and highlights the needs for researchers to propose automated fault localization tools for DL software deployment. Similarly, proactive alerting techniques can be proposed to inform developers about potential errors during the deployment process. 
However, monitoring and troubleshooting the deployment process is quite difficult, because of myriad potential issues, including hardware and software failures, misconfigurations, input data, even simply unrealistic user expectations, etc. Therefore, we encourage researchers to conduct a systematic study to characterize the major types and root causes of faults occurring during deployment of DL software before developing the aforementioned automated tools.
\textbf{(2) Automated configuration.} In our taxonomy, many challenges are related to configuration (e.g., \emph{Specification of model information} and \emph{Configuration of environment variables}). This observation motivates researchers to propose automated configuration techniques to simplify some deployment tasks for developers, especially non-experts. In addition, automated configuration checkers can be proposed to detect and diagnose misconfigurations, based on analyzing the configuration logic, requirements, and constraints. 
\textbf{(3) Implications for other communities.} Our results reveal some emerging needs of developers and can provide implications for other research communities, such as \emph{systems} and \emph{AI}. For example, some developers call for more quantization options (see \emph{Model quantization}) in model conversion. To help improve current frameworks, researchers from the AI community should propose more effective and efficient techniques for model quantization. In addition, to update models on mobile devices (see \emph{Model Update}), system researchers need to propose effective techniques to support model update (i.e., re-training) on the devices with limited computation power.

\subsection{Developers}
\textbf{(1) Targeted learning of required skills.} DL software deployment lies in the interaction between DL and SE. Therefore, DL software deployment requires developers with  solid knowledge of both fields, making this task quite challenging. 
Our taxonomy can serve as a checklist for developers with varying backgrounds, motivating the developers to learn necessary knowledge before really deploying DL software. For instance, an Android developer needs to learn necessary knowledge about DL before deploying DL software to mobile devices. Otherwise, she may fail in the specification of information about DL models (see \emph{Specification of model information}) trained by DL developers or data scientists. Similarly, when a DL developer who is not skillful in JavaScript deploys DL models on browsers, she may directly load models from local file systems due to the misunderstanding of browsers' local storage (see Section~\ref{un_brow}).
\textbf{(2) Avoiding common pitfalls.} Our study identifies some common pitfalls in DL software deployment. Developers should pay attention to these pitfalls and avoid them accordingly. For instance, when deploying DL software to target platforms,  developers should remember to migrate the pre-processing code and pay attention to version compatibility. \textbf{(3) Better project management.} Our taxonomy presents the distribution of  different categories, indicating which challenges developers have encountered more. In a project that involves DL software deployment, the project manager can use our taxonomy to assign a task where developers often have challenges (e.g., model conversion) to a more knowledgeable developer.

\subsection{Framework Vendors}
\textbf{(1) Improving the usability of documentation.}
As shown in our results, many developers even have difficulty in the entire procedure of deployment (i.e., how to deploy DL software). For instance, such questions account for 13.4\% in mobile deployment. As described earlier, developers often complain about the poor documentation in these questions, revealing that the usability~\cite{icseAghajaniNVLMBL19} of relevant documentation should be improved. Specifically, DL framework vendors can provide better detailed documentation and tutorials for developers' reference. In addition, confused information organization, such as hiding explanations of important concepts behind hyperlinks (see Section~\ref{un_brow}), may result in developers' misuse and thus should be avoided. 
\textbf{(2) Improving the completeness of documentation.} The prevalence of the  ``\emph{Conceptual questions}'' category suggests that framework vendors should improve the completeness~\cite{jssZhiGSGSR15,icseAghajaniNVLMBL19} of the documentation, especially considering that DL software deployment requires a wide set of background knowledge and skills. Indeed, basic information that might look clear from the vendors' perspective may not be easy to digest by the users (i.e., the developers)~\cite{icseAghajaniNVLMBL19}. The vendors should involve the users in the review of documentation, in order to supplement necessary explanations of basic knowledge in the documentation. This way might help in minimizing developers' learning curve and avoiding misunderstanding. 
\textbf{(3) Improving the design of APIs.} The quality of APIs heavily influences the development experience of developers and even correlates with the success of applications that make use of the APIs~\cite{VasquezBBPOP13}. Our study reveals some APIs' issues that need the attention of DL framework vendors. For one functionality, framework vendors may provide similar APIs for various options (see \emph{Selection/usage of APIs}), making some developers confused in practice. To mitigate this issue, framework vendors should better distinguish these APIs and clarify their use cases more clearly.
\textbf{(4) Improving functionalities as needed.} We observe that many developers suffer from conversion and export of unsupported models in the deployment process. For instance, in mobile deployment, 6.1\% of questions are about this challenge. Since it is impractical for framework vendors to support all possible operators at once, we suggest that framework vendors can mine SO and GitHub to collect related issues reported by developers and then first meet those most urgent operators and models. 


\section{Threats to Validity}\label{threats}
In this section, we discuss some threats to the validity of our study.

\textbf{Selection of tags and keywords.} Our automated identification of relevant questions is based on pre-selected tags and keyword-matching mechanisms, and thus may result in potential research bias. For tags, we mainly follow previous related work~\cite{sigsoftIslamNPR19,corrabstaxonomy,GuoCXMHLLZL19} to determine which ones to choose. Moreover, all tags that we use are about popular frameworks or platforms, promising the representativeness of the questions used in this study. However, it is still possible that in other contexts developers discuss issues that we do not encounter. In addition, the keyword-matching identification may result in the retrieval of false positives and the loss of posts that do not contain explicit keywords. The false positives are discarded during our manual examination of data for R3, so false positives do not affect the precision of our final taxonomy. However, due to the large amount of data used for R1 and R2, we do not employ similar manual examination to remove false positives for the two research questions, and thus may cause the results of R1 and R2 to be biased. Furthermore, although our identified posts with explicit keywords are more representative compared to the implicit posts, loss of implicit posts may introduce bias in our results.

\textbf{Selection of data source.} Similar to previous studies~\cite{esemAhmedB18,sigsoftBagherzadehK19,AlshangitiSMLY19,eseRosenS16,issredeep19,jcstYangLXWS16,chiprivacy,YILINGBUILD}, our work uses SO as the only data source to study the challenges that developers encounter. As a result, we may overlook valuable insights from other sources. In future work, we plan to extend our study to diverse data sources and conduct in-depth interviews with researchers and practitioners to further validate our results. However, since SO contains both novices' and experts' posts~\cite{isstaZhangCCXZ18}, we believe that our results are still valid. 

\textbf{Subjectivity of inspection.} The manual analysis in this study presents threats to the validity of our taxonomy. To minimize this threat, two authors are involved in inspecting cases and finally reach agreement with the help of three experienced arbitrators through discussions. The inter-rater agreement is relatively high, demonstrating the reliability of the coding schema and procedure.

\section{Related Work}\label{related}
In this section, we summarize related studies to  position our work within the literature.

\textbf{Challenges that ML/DL poses for SE.}
The rapid development of ML technologies poses new challenges for software developers. To characterize these challenges, Thung et al.~\cite{issreThungWLJ12} collect and analyze bugs in ML systems to study bug severity, efforts needed to fix bugs, and bug impacts. 
Alshangiti et al.~\cite{AlshangitiSMLY19} demonstrate that ML questions are more difficult to answer than other questions on SO and that model deployment is most challenging across all ML phases. In addition, Alshangiti et al.~\cite{AlshangitiSMLY19} find that DL related topics are most popular among the ML related questions. In recent years, several studies focus on the challenges in DL. By inspecting DL related posts on SO, Zhang et al.~\cite{issredeep19} find that program crashes, model deployment, and implementation questions are the top three most frequently asked questions. 
Besides, several studies characterize faults in software that makes use of DL frameworks. Zhang et al.~\cite{isstaZhangCCXZ18} collect bugs in TF programs from SO and GitHub. By manual examination, Zhang et al.~\cite{isstaZhangCCXZ18} categorize the symptoms and root causes of these bugs and propose strategies to detect and locate DL bugs. Following this work, Islam et al.~\cite{sigsoftIslamNPR19} and Humbatova et al.~\cite{corrabstaxonomy} extend the scope to the bugs in programs written based on the top five popular DL frameworks to present more comprehensive results.
Inspired by these previous studies, we also aim to investigate the challenges that DL poses for SE. However, different from these previous studies, our study focuses on the deployment process of DL software.

\textbf{DL software deployment.}
To make DL software really accessible for users, developers need to deploy DL software to different platforms according to various application scenarios. A popular way is to deploy DL software to server/cloud platforms, and then the DL functionality can be accessed as services. For this deployment way, Cummaudo et al.~\cite{corrabseduc} analyze pain points that developers face when using these services. In other words, Cummaudo et al.~\cite{corrabseduc} focus on challenges that occur after the deployment of DL software. Different from this work, our study focuses on  challenges in the deployment process. In addition, mobile devices have created great opportunities for DL software. Researchers have built numerous DL software applications on mobile devices~\cite{imwutXuQMHL18,hucRaduLBMMK16,hucMittalYGK16} and proposed various optimization techniques (e.g., model compression~\cite{cvprWuLWHC16,mobisysLiuLZNLD18} and cloud offloading~\cite{asplosKangHGRMMT17,mobicomXuZLLL18}) for deploying DL software to mobile platforms. To bridge the knowledge gap between research and practice, Xu et al.~\cite{wwwXuLLLLL19} conduct the first empirical study on large-scale Android apps to demystify how DL techniques are adopted in the wild. In addition, in recent years, various JavaScript-based DL frameworks have been published to enable DL-powered Web applications in browsers. To investigate what and how well we can do with these frameworks, Ma et al.~\cite{wwwMaXZTL19} select seven JavaScript-based frameworks and measure their performance gap when running different DL tasks on Chrome. The findings show that DL in browsers is still at dawn. Recently, Guo et al.~\cite{GuoCXMHLLZL19} put attention on DL software deployment across different platforms, and investigate the performance gap when the trained DL models are migrated from PC to mobile devices and Web browsers. The findings unveil that the deployment still suffers from compatibility and reliability issues. Despite these previous efforts,  specific challenges in deploying DL software are still under-investigated and thus our study aims to fill this knowledge gap.
\section{Conclusion}\label{conclusion}
Based on SO posts related to DL software deployment, in this paper, we have presented study findings to show that this task is becoming increasingly popular among software engineers. Furthermore, our findings demonstrate that DL software deployment is more challenging than other aspects of DL software and even other challenging topics in SE such as big data and concurrency, motivating us to identify the specific challenges behind DL software deployment. To this end, we manually inspect 769 sampled SO posts to derive a taxonomy of 72 challenges encountered by developers in DL software deployment. Finally, we qualitatively discuss our findings and infer implications for researchers, developers, and DL framework vendors, with the goal of highlighting good practices and valuable research avenues in deploying DL software.

\balance

\begin{acks}
This work was partially supported by the Key-Area Research and Development Program of Guangdong Province under the grant number 2020B010164002, the National Natural Science Foundation of China under the grant number 61725201, and the Beijing Outstanding Young Scientist Program under the grant number BJJWZYJH01201910001004. Haoyu Wang's work was supported by the National Natural Science Foundation of China under the grant number 61702045.
\end{acks}

\balance

\bibliographystyle{ACM-Reference-Format}
\bibliography{dldeploybib}

\end{document}